\documentclass[12pt, fleqn]{article}
\usepackage[cp1251]{inputenc}
\usepackage{latexsym,amsfonts,amssymb}
\usepackage{graphicx}

\usepackage{amsbsy}
\usepackage{amsmath}
\usepackage{epsf}

\newtheorem{theo}{Theorem}
\newcommand{\bt}{\begin{theo}}
\newcommand{\et}{\end{theo}}
\newcommand{\bd}{\begin{displaymath}}
\newcommand{\ed}{\end{displaymath}}

\newcommand{\be} {\begin{equation}}
\newcommand{\ee} {\end{equation}}
\newcommand{\ba} {\begin{array}}
\newcommand{\ea} {\end{array}}

\newcommand{\p} {\partial}

\newcommand{\al} {\alpha}
\newcommand{\lbd} {\lambda}

\begin{document}
\begin{titlepage}

\normalsize
 \begin{center}
 {\Large \bf Conditional symmetries
 and exact solutions \\ of
 the diffusive Lotka-Volterra system }\\
 \medskip

{\bf Roman Cherniha$^{\dag,\ddag}$} \footnote{\small e-mail:
cherniha@imath.kiev.ua}
 {\bf and  Vasyl' Davydovych$^{\dag}$}
 \footnote{\small e-mail: davydovych@imath.kiev.ua }
 \\
{\it  $^\dag$~Institute of Mathematics, Ukrainian National Academy
of Sciences,\\
 Tereshchenkivs'ka Street 3, Kyiv 01601, Ukraine}\\
 {\it $^\ddag$~Department of Mathematics, National University `Kyiv Mohyla Academy'\\
2, Skovoroda Street,   Kyiv 01601, Ukraine}\\
\end{center}

\renewcommand{\abstractname}{Abstract}
\begin{abstract}
 Q-conditional symmetries of
 the classical  Lotka-Volterra system in the  case of one space
 variable are completely described and a set of such symmetries
 in explicit form is constructed.
The relevant non-Lie ans\"atze  to reduce the classical
Lotka-Volterra systems with correctly-specified coefficients  to ODE
systems and examples of  new exact solutions  are found. A possible
biological  interpretation  of some exact  solutions  is presented.
\end{abstract}


\end{titlepage}

\begin{center}
{\bf 1.Introduction.}
\end{center}

Since 1952 when A.C. Turing published the remarkable paper
\cite{turing},  which  a revolutionary idea about mechanism of
morphogenesis (the development of structures in an organism during
the life) has been proposed in,
 nonlinear reaction-diffusion  
systems of the form  
\be\label{1}\ba{l} 
\lbd_1 u_t =  u_{xx}+F(u,v),\\  
\lbd_2 v_t = v_{xx}+G(u,v),  
\ea\ee  
have been extensively studied by means of different mathematical
methods, including group-theoretical methods (see
\cite{ch-king,ch-king2,niki-05} and the papers cited therein).
In system (\ref{1}),  $F$ and $G$ are arbitrary smooth functions,  
$u=u(t,x)$ and  $v=v(t,x)$ are unknown functions of the  variables  
$t, x $, while  the  subscript $t$ and $x$
denotes differentiation with respect to this variable.  
Notably  nonlinear system (\ref{1}) generalizes many well-known nonlinear  
second-order models  used to describe  
various processes in physics \cite{ames},   
biology \cite{mur2, britton} and ecology \cite{okubo}.

In the present paper, we shall consider the diffusive Lotka-Volterra
 (DLV) system
\be\label{2}\ba{l}
\lbd_1 u_t =  u_{xx}+u(a_1+b_1u+c_1v),\\ 
\lbd_2 v_t = v_{xx}+ v(a_2+b_2u+c_2v),  
\ea\ee  
which is the most common particular case of reaction-diffusion (RD)
system (\ref{1}). System (\ref{1}) is the simplest generalization of
the classical Lotka-Volterra  system that takes into account the
diffusion process for interacting species (see terms  $u_{xx}$ and
$v_{xx}$). Nevertheless the classical Lotka-Volterra  system  was
introduced by A.J.Lotka and V.Volterra more than 80 years ago, its
natural generalization (\ref{2}) is still  studied because this is
one of the most important mathematical models.   Lie symmetries of
(\ref{2}) have been completely described in \cite{ch-du-04} (note
those can be extracted from more general results presented in
\cite{ch-king,ch-king2}).

The problem
 of construction of $Q$-conditional symmetries (non-classical symmetries)
  for (\ref{1})  is  still not
  solved even in the case of DLV system (\ref{2}).
  Moreover, to  our best knowledge, there are only a few papers
  devoted to the search of conditional symmetries
  of     {\it systems of PDEs}. Notably, some general results about
  $Q$-conditional symmetries of RD systems with power diffusivities
  of the form
 \be\label{1**}\ba{l}
 u_t=(u^ku_x)_x+F(u,v),\\
v_t=(v^lv_x)_x+ G(u,v)
 \ea\ee
 have been obtained in the recent paper \cite{ch-pli-08}.
 However, the results obtained in \cite{ch-pli-08} cannot be adopted
 for any system of the form (\ref{1}) because the case $l=k=0$
 is a very special and wasn't studied therein.

 It should be noted that there are many papers  devoted to the construction of  such symmetries
 for the  {\it scalar} non-linear reaction-diffusion (RD) equations of the form
 \cite{fss,   nucci1, cla, a-b-h, a-h, pucci-sac2000}
 \be\label{2**}
 U_t=\left[D(U)U_x\right]_x + F(U)
\ee and (\ref{2**}) with the  convective term   $B(U)U_x$ (here
$B(U), D(U)$ and $F(U)$ are  arbitrary smooth functions) \cite{
ch-se-98,che-2006,ch-pliu-2006}.

It is well-known that conditional symmetries can be applied for
finding exact solutions of the relevant equations, which are not
obtainable by the classical Lie method. Moreover the solutions
obtained in such a way may have   a  physical or biological
interpretation (see, e.g.,  examples in  \cite{
che-2006,ch-pliu-2006,dix-cla, broad-2004}) what is of fundamental
importance.

  The paper is organized as follows.
  In Section 2, we present  two  definitions of $Q$-conditional
  symmetry in the case of RD system (\ref{2}) and show
  how they are connected with non-classical symmetry.
In   Section 3,  we present a complete description of
     $Q$-conditional symmetries of
   the DLV system (\ref{2}), i.e. the system of the determining equations for constructing $Q$-conditional
symmetries of system (\ref{1}) is derived and analyzed. Here the
main theorems presenting these symmetries in explicit form are
proved. In Section 4, the $Q$-conditional symmetries obtained are
applied to  reduce
 the corresponding DLV systems
to the systems of ordinary differential equations (ODEs) and
constructing exact solutions. The properties of an exact  solution
are examined with the aim to provide the relevant interpretation for
population dynamics.

 Finally, we present some conclusions.

\begin{center}
{\textbf{2. Definitions of conditional symmetry for systems of
PDEs.}}
\end{center}

Here we  present  new  definitions of $Q$-conditional symmetry which
naturally arise for {\it systems} of PDEs. To avoid possible
difficulties that can occur in the case of arbitrary system of PDEs,
we restrict ourself on the RD systems of the form (\ref{1}).

It is well-known that to find Lie invariance  operators, one needs
to consider   system  
(\ref{1}) as the manifold ${\cal{M}}=\{S_1=0,S_2=0 \}$  where
\be\label{2-1}  \ba{l}
 S_1 \equiv \lbd_1 u_t -  u_{xx}- F(u,v)=\,0,\\  
S_2 \equiv \lbd_2 v_t -  v_{xx}- G(u,v)=\,0,
\ea\ee  
in the prolonged space of the  variables:  
$t, x, u, v, u_t, v_t$,$ u_{x}, v_{x}, u_{xx}, v_{xx}, u_{xt}, v_{xt}, u_{tt}, v_{tt}.$ 
According to the definition, system  (\ref{1}) is invariant under the transformations generated by the  
infinitesimal operator  
\be\label{2-2}  
Q = \xi^0 (t, x, u, v)\p_{t} + \xi^1 (t, x, u, v)\p_{x} +  
 \eta^1(t, x, u, v)\p_{u}+\eta^2(t, x, u, v)\p_{v},  \ee  
if the following invariance conditions are satisfied:  
\be\label{2-3}  
\ba{l}  
\mbox{\raisebox{-1.1ex}{$\stackrel{\displaystyle  
Q}{\scriptstyle 2}$}} S_1  
 \equiv  \mbox{\raisebox{-1.1ex}{$\stackrel{\displaystyle  
Q}{\scriptstyle 2}$}}  
(\lbd_1 u_t -  u_{xx}- F(u,v) )  
\Big\vert_{\cal{M}}=0, \\[0.3cm]  
\mbox{\raisebox{-1.1ex}{$\stackrel{\displaystyle  
Q}{\scriptstyle 2}$}} S_2  
 \equiv  \mbox{\raisebox{-1.1ex}{$\stackrel{\displaystyle  
Q}{\scriptstyle 2}$}}  
(\lbd_2 v_t -  v_{xx}- G(u,v) )  
\Big\vert_{\cal{M}}=0.  
\ea  
\ee  
The operator $ \mbox{\raisebox{-1.1ex}{$\stackrel{\displaystyle  
Q}{\scriptstyle 2}$}} $  
is the second  
 prolongation of the operator $Q$, i.e.  
\be\label{2-4}  
\mbox{\raisebox{-1.1ex}{$\stackrel{\displaystyle  
Q}{\scriptstyle 2}$}}  
 = Q + \rho_t^1\frac{\partial}{\partial u_{t}}+
 \rho_t^2\frac{\partial}{\partial v_{t}}+  
\rho^1_x\frac{\partial}{\partial u_{x}}+ \rho^2_x\frac{\partial}{\partial v_{x}}  
+\sigma_{xx}^1\frac{\partial}{\partial u_{xx}}  
+\sigma_{xx}^2\frac{\partial}{\partial  v_{xx}},  
\ee  
where the coefficients $\rho$ and $\sigma$ with relevant subscripts  
are expressed  via the functions $\xi^0, \xi^1, \eta^1$ and $\eta^2$
by well-known formulae (see, e.g., \cite{fss, olv, b-k}).

The crucial idea used for introducing the notion of $Q$-conditional
symmetry (non-classical symmetry) is to change  the manifold
${\cal{M}}$, namely: the operator $Q$ is used to reduce ${\cal{M}}$.
It can be noted that there are two essentially different
possibilities to realize this idea in the case of system (\ref{1}).

\noindent {\bf Definition 1.} Operator (\ref{2-2}) is called the
$Q$-conditional symmetry of the first type  for the RD system
(\ref{1}) if  the following invariance conditions are satisfied:
\be\label{2-5}  
\ba{l}  
\mbox{\raisebox{-1.1ex}{$\stackrel{\displaystyle  
Q}{\scriptstyle 2}$}} S_1  
 \equiv  \mbox{\raisebox{-1.1ex}{$\stackrel{\displaystyle  
Q}{\scriptstyle 2}$}}  
(\lbd_1 u_t -  u_{xx}- F(u,v) )  
\Big\vert_{{\cal{M}}_1}=0, \\[0.3cm]  
\mbox{\raisebox{-1.1ex}{$\stackrel{\displaystyle  
Q}{\scriptstyle 2}$}} S_2  
 \equiv  \mbox{\raisebox{-1.1ex}{$\stackrel{\displaystyle  
Q}{\scriptstyle 2}$}}  
(\lbd_2 v_t -  v_{xx}- G(u,v) )  
\Big\vert_{{\cal{M}}_1}=0,  
\ea  
\ee  
where the manifold ${\cal{M}}_1$ is either $\{S_1=0,S_2=0, Q(u)=0
\}$ or $\{S_1=0,S_2=0, Q(v)=0 \}$.

\noindent \textbf{Definition 2.} Operator (\ref{2-2}) is called the
$Q$-conditional symmetry of the second  type, i.e., the standard
$Q$-conditional symmetry (non-classical symmetry) for the RD system
(\ref{1}) if  the following invariance conditions are satisfied:
\be\label{2-6}  
\ba{l}  
\mbox{\raisebox{-1.1ex}{$\stackrel{\displaystyle  
Q}{\scriptstyle 2}$}} S_1  
 \equiv  \mbox{\raisebox{-1.1ex}{$\stackrel{\displaystyle  
Q}{\scriptstyle 2}$}}  
(\lbd_1 u_t -  u_{xx}- F(u,v) )  
\Big\vert_{{\cal{M}}_2}=0, \\[0.3cm]  
\mbox{\raisebox{-1.1ex}{$\stackrel{\displaystyle  
Q}{\scriptstyle 2}$}} S_2  
 \equiv  \mbox{\raisebox{-1.1ex}{$\stackrel{\displaystyle  
Q}{\scriptstyle 2}$}}  
(\lbd_2 v_t -  v_{xx}- G(u,v) )  
\Big\vert_{{\cal{M}}_2}=0,  
\ea  
\ee  
where the manifold ${\cal{M}}_2$ = $\{S_1=0,S_2=0, Q(u)=0,
Q(v)=0\}$.

It is easily seen that ${\cal{M}}_2 \subset {\cal{M}}_1 \subset
{\cal{M}}$, hence, each Lie symmetry is automatically a
$Q$-conditional symmetry of the first and second  type, while
$Q$-conditional symmetry of the first type is one of the  second
type. From the formal point of view is enough to find all the
$Q$-conditional symmetry of the second type. Having the full list of
$Q$-conditional symmetries of the second  type, one may simply check
which of them is Lie symmetry or/and  $Q$-conditional symmetry of
the first type.

On the other hand,
 to construct   $Q$-conditional symmetries of both types for  a
system of PDEs, one needs to solve new nonlinear system, so called
system of determining equations, which usually is much more
complicated than one for searching Lie symmetries. This problem
arises even in the case of linear single PDE and it was the reason
why G.Bluman and J.Cole in their pioneering work \cite{bl-c} were
unable to describe all $Q$-conditional symmetries in explicit form
even for the linear heat equation. Thus, both definition are
important from theoretical and practical point of view.

It should be noted that Definition 2 was only used in papers
\cite{ch-pli-08, murata-06, ch-se-03}  devoted to the search
$Q$-conditional symmetries for the systems of PDEs. Moreover, to our
best knowledge, nobody has noted that a hierarchy of conditional
symmetry operators can be defined for systems involving, say, $m$
PDEs. In fact, different definitions can be formulated for such
systems in quite similar way to Definitions 1 and 2 (see
\cite{ch-2010} for details).

\begin{center}
{\textbf{3. Conditional symmetries of the DLV system (\ref{2}) }}
\end{center}

First of all,  we construct the system of determining equations(DEs)
to construct $Q$-conditional symmetries  of the second type
(nonclassical symmetries) of system (\ref{2}). The most general form
of such operators is the first-order operator \be \label{3*}
 {Q}= \
 \xi^0(t,x,u,v)\partial_t+\xi^1(t,x,u,v)\partial_x+\eta^1(t,x,u,v)\partial_u+\eta^2(t,x,u,v)\partial_v,
\ee where   the  functions $\xi^i(t,x,u,v)$ and $\eta^k(t,x,u,v)$
should be determined from the relevant system of
 DEs . In the case $\xi^0(t,x,u,v)\not=0$, this
system  can be reduced to that with $\xi^0(t,x,u,v)=1$
\cite{ch-2010} so that we are looking for the operators \be
\label{3}
 {Q}= \
 \partial_t+\xi(t,x,u,v)\partial_x+\eta^1(t,x,u,v)\partial_u+\eta^2(t,x,u,v)\partial_v.
\ee

Note we examine system (\ref{2}) only in those case when one is a
real system of coupled equations, i.e., $b^2_2+c^2_1\neq0$, and
contain non-linear equations (otherwise the system is rather
artificial).

Let us apply Definition 2 to construct the system of DEs for finding
operator (\ref{3}). According to the definition the following
invariance conditions must be satisfied:
\be\label{3-4}\ba{l}  
  \mbox{\raisebox{-1.1ex}{$\stackrel{\displaystyle  
Q}{\scriptstyle 2}$}}  
\left( \lbd_1 u_t =  u_{xx}+u(a_1+b_1u+c_1v)
\right)\Big\vert_{{\cal{M}}}
=0,\\
\mbox{\raisebox{-1.1ex}{$\stackrel{\displaystyle  
Q}{\scriptstyle 2}$}}  
\left( \lbd_2 v_t = v_{xx}+ v(a_2+b_2u+c_2v)
\right)\Big\vert_{{\cal{M}}} =0, \ea \ee
 where the manifold
\be\label{3-4*}{\cal{M}}= \{S_1=0,\ S_2=0, \ u_t+ \xi u_x=\eta^1,\
v_t+\xi v_x=\eta^2 \} \ee (here the left-hand-sides of (\ref{2}) are
denoted as $S_1$ and $S_2$) while
\[\ba{l} \medskip
\mbox{\raisebox{-1.1ex}{$\stackrel{\displaystyle Q}{\scriptstyle
2}$}}
=Q+\rho^1_t\frac{\p}{\p u_t}+\rho^2_t\frac{\p}{\p v_t}+\rho^1_x\frac{\p}{\p u_x}+\rho^2_x\frac{\p}{\p v_x} \\
\qquad +\sigma^1_{tx}\frac{\p}{\p u_{tx}}+\sigma^2_{tx}\frac{\p}{\p
v_{tx}} +\sigma^1_{tt}\frac{\p}{\p u_{tt}}+\sigma^2_{tt}\frac{\p}{\p
v_{tt}} +\sigma^1_{xx}\frac{\p}{\p u_{xx}}+\sigma^2_{xx}\frac{\p}{\p
v_{xx}} \ea \]
is the second order  
 prolongation of the operator $Q$ and its coefficients  
are expressed  via the functions $\xi,  \eta^1 $, and $ \eta^2$ by
well-known formulae (see, e.g.,\cite{fss, olv, b-k}).

Now we apply the rather standard procedure for obtaining system of
DEs, using the  invariance conditions (\ref{3-4}). From the formal
point of view, the procedure is the same as for Lie symmetry search,
however, four (not two !) different derivatives, say $u_{xx},
v_{xx}, u_t $ and $v_t$, can be excluded using the manifold
${\cal{M}}$. After   straightforward calculations, one arrives at
the nonlinear system of DEs
 \be\label{6}\xi_{uu}=\xi_{vv}=\xi_{uv}=0 ,\ee
\be\label{7}\eta^1_{vv}=0  ,\ee \be\label{8}\eta^2_{uu}=0  ,\ee
\be\label{9}\ 2\lambda_1\xi\xi_u +\eta^1_{uu}-2\xi_{xu}=0,\ee
\be\label{9*}\ 2\lambda_2\xi\xi_v +\eta^2_{vv}-2\xi_{xv}=0,\ee
\be\label{10}(\lambda_1+\lambda_2)\xi\xi_v+2\eta^1_{uv}-2\xi_{xv}=0
,\ee
\be\label{11}(\lambda_1+\lambda_2)\xi\xi_u+2\eta^2_{uv}-2\xi_{xu}=0
,\ee
\be\label{12}(\lambda_1-\lambda_2)\xi\eta^1_v+2\eta^1_{xv}+2u(a_1+b_1u+c_1v)\xi_v
-2\lambda_1\xi_v\eta^1=0  ,\ee \be\label{13}
(\lambda_2-\lambda_1)\xi\eta^2_u+2\eta^2_{xu}+2v(a_2+b_2u+c_2v)\xi_u
-2\lambda_2\xi_u\eta^2=0 ,\ee
\be\label{15}\ba{l} 
\lambda_1(2\xi_u\eta^1-\xi_t-\xi_v\eta^2-2\xi\xi_x)+\lambda_2\xi_v\eta^2-3\xi_uu(a_1+b_1u+c_1v)-\\  
-\xi_vv(a_2+b_2u+c_2v)-2\eta^1_{xu}+\xi_{xx}=0,  
\ea\ee  
\be\label{16}\ba{l} 
\lambda_2(2\xi_v\eta^2-\xi_t-\xi_u\eta^1-2\xi\xi_x)+\lambda_1\xi_u\eta^1-3\xi_vv(a_2+b_2u+c_2v)-\\  
-\xi_uu(a_1+b_1u+c_1v)-2\eta^2_{xv}+\xi_{xx}=0,  
\ea\ee  
\be\label{17a}\ba{l} 
\lambda_1(\eta^1_t+\eta^2\eta^1_v+2\xi_x\eta^1)-\lambda_2\eta^2\eta^1_v-\eta_1(a_1+2b_1u+c_1v)-c_1\eta_2v+\\  
+\eta^1_uu(a_1+b_1u+c_1v)-2\xi_xu(a_1+b_1u+c_1v)+\eta^1_vv(a_2+b_2u+c_2v)-\eta^1_{xx}=0,  
\ea\ee  
\be\label{17}\ba{l} 
\lambda_2(\eta^2_t+\eta^2\eta^2_u+2\xi_x\eta^2)-\lambda_1\eta^1\eta^2_u-\eta_2(a_2+b_2u+2c_2v)-b_2\eta_1v+\\  
+\eta^2_uu(a_1+b_1u+c_1v)-2\xi_xv(a_2+b_2u+c_2v)+\eta^2_vv(a_2+b_2u+c_2v)-\eta^2_{xx}=0.  
\ea\ee  

It turns out,  the functions $ \eta^1$ and $ \eta^2$ can be at
maximum linear functions with respect to  $u$ and $v$. In fact, the
differential consequences of   (\ref{10}) and  (\ref{11}) with
respect to these variables  lead to the expressions
\[(\lambda_1+\lambda_2)\xi^2_v=0,$ $(\lambda_1+\lambda_2)\xi^2_u=0 \]
so that  $\xi_u=\xi_v=0$. Having $\xi=\xi(t,x)$ equations
(\ref{7})--(\ref{11}) can be easily solved and one arrives at
\be\label{18}\ba{l}
\eta^1=q^1(t,x)v+r^1(t,x)u+p^1(t,x)\\
\eta^2=q^2(t,x)u+r^2(t,x)v+p^2(t,x), \ea\ee hence, the most general
form of operator  (\ref{3}) for system (\ref{2}) is as follows
\be\label{19}{Q}=\partial_t+\xi\partial_x+(q^1v+r^1u+p^1)\partial_u+(q^2u+r^2v+p^2)\partial_v,\ee
where   the functions $q^k, r^k, p^k \ (k=1,2)$  should be found
from other equations. Substituting  (\ref{18}) and $\xi_u=\xi_v=0$
into equations (\ref{12})--(\ref{17}), those  can be splitted with
respect to $u, v, u^2, v^2, uv$. Finally, one obtains the system of
DEs

\be\label{20}(c_1-c_2)q^1=0 ,\ee \be\label{21}(b_1-b_2)q^2=0 ,\ee
\be\label{22}c_1q^2+b_1(r^1+2\xi_x)=0 ,\ee
\be\label{23}b_2q^1+c_2(r^2+2\xi_x)=0 ,\ee
\be\label{24}(2b_1-b_2)q^1+c_1(r^2+2\xi_x)=0 ,\ee
\be\label{25}(2c_2-c_1)q^2+b_2(r^1+2\xi_x)=0 ,\ee
\be\label{26}(\lambda_1-\lambda_2)\xi q^1+2q^1_x=0 ,\ee
\be\label{27}(\lambda_2-\lambda_1)\xi q^2+2q^2_x=0 ,\ee
\be\label{28}\lambda_1(\xi_t+2\xi\xi_x)+2r^1_x-\xi_{xx}=0 ,\ee
\be\label{29}\lambda_2(\xi_t+2\xi\xi_x)+2r^2_x-\xi_{xx}=0 ,\ee
\be\label{30}\lambda_1(r^1_t+2r^1\xi_x)+(\lambda_1-\lambda_2)q^1q^2-c_1p^2-2b_1p^1-2a_1\xi_x-r^1_{xx}=0
,\ee
\be\label{31}\lambda_2(r^2_t+2r^2\xi_x)+(\lambda_2-\lambda_1)q^1q^2-b_2p^1-2c_2p^2-2a_2\xi_x-r^2_{xx}=0
,\ee
\be\label{32}\lambda_1(q^1_t+2q^1\xi_x)+(\lambda_1-\lambda_2)q^1r^2-(a_1-a_2)q^1-c_1p^1-q^1_{xx}=0
,\ee
\be\label{33}\lambda_2(q^2_t+2q^2\xi_x)+(\lambda_2-\lambda_1)q^2r^1+(a_1-a_2)q^2-b_2p^2-q^2_{xx}=0
,\ee
\be\label{34}\lambda_1(p^1_t+2p^1\xi_x)+(\lambda_1-\lambda_2)q^1p^2-a_1p^1-p^1_{xx}=0
,\ee
\be\label{35}\lambda_2(p^2_t+2p^2\xi_x)+(\lambda_2-\lambda_1)q^2p^1-a_2p^2-p^2_{xx}=0
\ee to find $Q$-conditional symmetry operator (\ref{19}) of system
(\ref{2}).

\bt In the case $\lambda_1\neq\lambda_2$,  DLV system (\ref{2})  is
$Q$-conditionally invariant under operator (\ref{19}) if and only if
$b_1=b_2=b, c_1=c_2=c.$ Moreover, if  $bc=0, b^2+c^2\neq0$ then
system  (\ref{2}) and $Q$-conditional symmetries of the second type
(up to local transformations $u\rightarrow bu, \ v\rightarrow
\exp(\frac{a_2}{\lambda_2}t)v, \ b\neq0$  and  $u\rightarrow
\exp(\frac{a_1}{\lambda_1}t)v,\ cv\rightarrow u, \ c\neq0$ ) have
the forms \be\label{36} \ba{l}
\lbd_1 u_t= u_{xx}+u(a_1+u),\\ \lbd_2 v_t= v_{xx}+ v u, \\
\ea \ee and \be \label{37}{Q}=\p_t+\frac{2\al_1}{\lbd_1-\lbd_2} \,
\p_x+\Big(\exp
(\al_1x+\frac{\al^2_1}{\lbd_2}t)\big((\al_3+\al_4\exp(-\frac{a_1}{\lbd_2}t))u+\al_3a_1\big)+\al_2v\Big)\p_v,
\ee where  $\alpha_k, \ k= \overline{1,4}$
 are arbitrary constants with the restriction
$\alpha^2_3+\alpha^2_4\neq0.$

If  $bc\neq0$ and the additional restrictions  $q^1_x=q^2_x=0$ take
place then exactly three cases  (up to local transformations
$u\rightarrow bu, \ v\rightarrow cv $ and  $u\rightarrow v, \
v\rightarrow u $ ) exist when   system  (\ref{2}) admits
$Q$-conditional symmetry operators. They are listed as follows:
 \be\label{38}(i)
\ba{l}
 \quad \ \lbd_1 u_t = u_{xx}+u(a_1+u+v),\\
 \quad \ \lbd_2 v_t = v_{xx}+ v(a_2+u+v), \\
\ea \ee 
\be\label{39}\quad \
{Q_1}=(\lbd_1-\lbd_2)\p_t-(a_1v+a_2u+a_1a_2)(\p_u-\p_v), \
a^2_1+a^2_2 \neq 0,\ee \be\label{40} \quad \
{Q_2}=(\lbd_1-\lbd_2)\p_t+(a_1-a_2)u(\p_u-\p_v), \ a_1\neq a_2, \ee
\be\label{40***}\quad \ {Q}_3=(\lbd_1-\lbd_2)\p_t-(a_1-a_2)v(\p_u -
\p_v), \ a_1\neq a_2.\ee \be\label{41}(ii) \ba{l}
 \quad \ \lbd_1 u_t = u_{xx}+u(a+u+v),\\
 \quad \ \lbd_2 v_t = v_{xx}+ v(a+u+v), \\
\ea \ee \be\label{39*}\quad \
{Q_1}=(\lbd_1-\lbd_2)\p_t-a(v+u+a)(\p_u-\p_v), \ a\neq 0,\ee
 \be\label{42}
 \quad \
 {Q_2}=(\lbd_1-\lbd_2)t\p_t-(\lambda_1v+\lambda_2u)(\p_u-\p_v).
\ee \be\label{43}(iii) \ba{l}
 \quad \ \lbd_1 u_t = u_{xx}+u(a\lambda_1+u+v),\\
 \quad \ \lbd_2 v_t = v_{xx}+ v(a\lambda_2+u+v), \quad a\neq0,\\
\ea \ee \be\label{39**}\quad \ {Q_1}=(\lbd_1-\lbd_2)\p_t-a(\lbd_1v+
 \lbd_2u+a\lbd_1\lbd_2)(\p_u-\p_v), \ee
 \be\label{40*}
 \quad \  {Q_2}=\p_t+au(\p_u-\p_v), \ee
  \be\label{40**}
 \quad \  {Q_3}=\p_t-a v(\p_u-\p_v),\ee
\be\label{44}
 \quad
 \ {Q_4}=(e^{-at}-\alpha(\lbd_1-\lbd_2))\p_t+a\alpha(\lbd_1v+
 \lbd_2u+a\lbd_1\lbd_2)(\p_u-\p_v), \ \alpha\neq0.
\ee
 \et

\textbf{Proof.} Using equations (\ref{20})--(\ref{21}) from the
system of DEs, one notes that three cases can only take place:
 \textbf{1.1} $b_1\not=b_2$ and/or $ c_1\not=c_2;$
\textbf{1.2} $b_1=b_2=b\not=0,\ c_1=c_2=0,$ and \textbf{1.3}
$b_1=b_2=b\not=0,\ c_1=c_2=c\not=0.$ Note the fourth possible case $
b_1=b_2=0,\ c_1=c_2=c\neq0 $ is reduced  to the second case by the
renaming $u\rightarrow v, \rightarrow u $.

It turns out that case \textbf{1.1} produces   the restrictions
\be\label{45} q^1=0, \quad q^2=0,
 \ee which lead  only to Lie symmetry
operators of system (\ref{2}). Let us show this.

If $b_1\neq b_2$ and $ c_1\neq c_2$ then restrictions (\ref{45})
immediately  follow from (\ref{20})-- (\ref{21}). If  $b_1=b_2=b$
and $\ c_1\neq c_2$ then $q^1=0$ follows from (\ref{20}),
furthermore,  equations (\ref{22}) and (\ref{25}) lead to
$(c_2-c_1)q^2=0\Leftrightarrow q^2=0$ (the subcase $b_1\neq b_2, \
c_1=c_2=c$  leads to the same result).

Having restrictions (\ref{45}), we immediately obtain $c_1p^1=0, \
b_2p^2=0$ from (\ref{32}) and (\ref{33}). Obviously, if
$c_1b_2\neq0$, then \be\label{48} p^1=p^2=0. \ee If $c_1b_2=0$, say,
 $c_1=0, \ b_2\neq0$ ( subcase $c_1\neq0, \ b_2=0$ can be treated in the same way ) then $p^2=0$.   Simultaneously system
(\ref{2}) reduced to one with an autonomous equation, which is
nothing else but the Fisher equation
\be\label{46} \lbd_1 u_t =
u_{xx}+u(a_1+u), \quad a_1\neq0. \ee Now we assume that such system
is $Q$-conditionally invariant under an operator of the form
(\ref{19}). However, the Fisher equation doesn't admit any
$Q$-conditional symmetry \cite{ch-se-98} but only Lie symmetry
 \be\label{47}
{Q}=\p_t+\alpha\p_x, \ \alpha=const, \ee hence, setting $\xi=\alpha$
into (\ref{22}) -- (\ref{35}), we arrive at (\ref{48}). The special
value $a_1=0$ in (\ref{46}) leads to the additional Lie symmetry
$2t\p_t+x\p_x-2u\p_u$ but there aren't  $Q$-conditional symmetry
operators \cite{ch-se-98}. So, the restrictions (\ref{48}) are still
obtained.


Restrictions  (\ref{45}) and  (\ref{48}) simplify  essentially the
system of DEs (\ref{20}) -- (\ref{35}), which takes the form
\be\label{49}b_1(r^1+2\xi_x)=0 ,\ee \be\label{50}b_2(r^1+2\xi_x)=0
,\ee \be\label{51}c_1(r^2+2\xi_x)=0 ,\ee
\be\label{52}c_2(r^2+2\xi_x)=0 ,\ee
\be\label{53}\lambda_1(\xi_t+2\xi\xi_x)+2r^1_x-\xi_{xx}=0 ,\ee
\be\label{54}\lambda_2(\xi_t+2\xi\xi_x)+2r^2_x-\xi_{xx}=0 ,\ee
\be\label{55}\lambda_1(r^1_t+2r^1\xi_x)-2a_1\xi_x-r^1_{xx}=0 ,\ee
\be\label{56}\lambda_2(r^2_t+2r^2\xi_x)-2a_2\xi_x-r^2_{xx}=0.\ee Now
one may easily check that any solution of system
(\ref{49})--(\ref{56}) leads to a $Q$-conditional symmetry operator
which will be equivalent to Lie symmetry operator obtained in
\cite{ch-du-04}. Consider, for example, the most general case
$b_2c_1\neq0.$ Obviously,  (\ref{50}) and  (\ref{51}) with
$b_2c_1\neq0$ lead to \be\label{57}r^1=r^2=-2\xi_x.\ee So, having
(\ref{57}) and  $\lambda_1\neq\lambda_2,$  we obtain   the system
\be\label{58}\ba{l} 
 \xi_t+2\xi\xi_x=0,\\
 \xi_{xx}=0, \\
\ea \ee from (\ref{53}), (\ref{54}) and  (\ref{57}). Substituting
the general solution of (\ref{58})
\be\label{59}\xi(t,x)=\frac{x+\alpha_1}{2t+\alpha_2} \ee into
 (\ref{57}), we can solve equations
(\ref{55})-(\ref{56}). Finally, the operator
\be\label{59*}{Q}=\p_t+\frac{x+\alpha_1}{2t+\alpha_2}
\p_x-\frac{2}{2t+\alpha_2}(u\p_u+v\p_v)\ee is obtained  if
$a_1=a_2=0$. However, operator (\ref{59*}) is nothing else but a
linear combination of Lie symmetry operators of (\ref{2})
\cite{ch-du-04} (see table 1, case 1.) multiplied by
$\frac{1}{2t+\alpha_2}$. If $a_1^2+a_2^2\neq0$ then  operator
(\ref{47}) occurs, which  is  Lie symmetry operator. Thus, case
\textbf{1.1} is completely examined.

Consider  case \textbf{1.2}. Here system (\ref{2}) can be reduced to
one (\ref{36}) by the substitution $u\rightarrow bu, \ v\rightarrow
\exp(\frac{a_2}{\lambda_2}t)v,$  because  $c_1=c_2=0$. Since the
first equation of (\ref{36}) is the Fisher equation (\ref{46}) the
same approach can be used as above. Thus, using operator (\ref{47}),
we again substitute $\xi=\alpha$ into (\ref{22}) -- (\ref{35}), what
leads to the restrictions $q^1=p^1=r^1=0$ and  operator (\ref{19})
takes the form \be\label{61}{Q}=\partial_t+\alpha
\partial_x+(q^2u+r^2v+p^2)\partial_v.\ee
Simultaneously, the system of DEs (\ref{20}) -- (\ref{35}) reduces
to \be\label{62}r^2_x=r^2_t=0,\ee
\be\label{63}(\lambda_2-\lambda_1)\xi q^2+2q^2_x=0 ,\ee
\be\label{64}\lambda_2q^2_t+a_1q^2-p^2-q^2_{xx}=0 ,\ee
\be\label{65}\lambda_2p^2_t-p^2_{xx}=0 .\ee The general solution of
this system  can be straightforwardly constructed and it reads as
follows
 \be\label{66}r^2=\alpha_2,\ee
\be\label{67}q^2=c(t)\exp\Big(\frac{\alpha}{2}(\lambda_1-\lambda_2)x\Big)
,\ee
\be\label{67*}p^2=\exp\Big(\frac{\alpha}{2}(\lambda_1-\lambda_2)x\Big)\Big(\lambda_2c'(t)
+a_1c(t)-\big(\frac{\alpha_1}{2}(\lambda_1-\lambda_2)\big)^2\Big)
,\ee where \be\label{68}
c(t)=\alpha_3\exp\Big(\frac{\alpha^2}{4\lambda_2}(\lambda_1-\lambda_2)^2t\Big)
+\alpha_4\exp\Big(\frac{1}{\lambda_2}\big(-a_1+\frac{\alpha^2}{4}(\lambda_1-\lambda_2)^2\big)t\Big),\ee
and  $\alpha_k, \ k=2,3,4$ are arbitrary constants. Finally,
introducing the notation
$\alpha=\frac{2}{\lambda_1-\lambda_2}\alpha_1$ and substituting
 the function
$r^2, q^2, \ p^2$  into  (\ref{61}), we arrive at the
$Q$-conditional symmetry operator (\ref{37}).

Consider  case \textbf{1.3}.
Here the DLV system (\ref{2}) is reduced to system (\ref{38}), i.e.
(\ref{2}) with $b=c=1$, by the substitution  $u\rightarrow bu, \
v\rightarrow cv$. Now we take into account the restrictions
$q^1_x=q^2_x=0,$ hence, the system of DEs reads as follows
\be\label{70}(\lambda_1-\lambda_2)\xi q^1=0 ,\ee
\be\label{71}(\lambda_2-\lambda_1)\xi q^2=0 ,\ee
\be\label{72}q^1+r^2+2\xi_x=0 ,\ee \be\label{73}q^2+r^1+2\xi_x=0
,\ee \be\label{74}\lambda_1(\xi_t+2\xi\xi_x)+2r^1_x-\xi_{xx}=0 ,\ee
\be\label{75}\lambda_2(\xi_t+2\xi\xi_x)+2r^2_x-\xi_{xx}=0 ,\ee
\be\label{76}\lambda_1(r^1_t+2r^1\xi_x)+(\lambda_1-\lambda_2)q^1q^2-p^2-2p^1-2a_1\xi_x-r^1_{xx}=0
,\ee
\be\label{77}\lambda_2(r^2_t+2r^2\xi_x)+(\lambda_2-\lambda_1)q^1q^2-p^1-2p^2-2a_2\xi_x-r^2_{xx}=0
,\ee
\be\label{78}\lambda_1(q^1_t+2q^1\xi_x)+(\lambda_1-\lambda_2)q^1r^2-(a_1-a_2)q^1-p^1=0
,\ee
\be\label{79}\lambda_2(q^2_t+2q^2\xi_x)+(\lambda_2-\lambda_1)q^2r^1+(a_1-a_2)q^2-p^2=0
,\ee
\be\label{80}\lambda_1(p^1_t+2p^1\xi_x)+(\lambda_1-\lambda_2)q^1p^2-a_1p^1-p^1_{xx}=0
,\ee
\be\label{81}\lambda_2(p^2_t+2p^2\xi_x)+(\lambda_2-\lambda_1)q^2p^1-a_2p^2-p^2_{xx}=0
.\ee

If  $q^1=q^2=0$ then we again obtain only Lie's operators (see the
case \textbf{1.1}). So, non-trivial results are obtainable only
under restriction $(q^1)^2+(q^2)^2\neq0$. Equations
(\ref{70})--(\ref{71}) under this restrictions  produce $\xi=0$,
hence, we obtain \be\label{82}q^1=-r^2=-\psi(t), \
q^2=-r^1=-\varphi(t)\ee from (\ref{72})--(\ref{75}) (here
$\varphi(t)$ and $ \psi(t)$ are arbitrary smooth functions at the
moment). Substituting (\ref{82}) into (\ref{78})--(\ref{79}), we
find
\be\label{83}\ba{l} \medskip
p^1=(a_1-a_2)\psi(t)+(\lambda_2-\lambda_1)\psi^2(t)-\lambda_1\psi'(t),\\
p^2=(a_2-a_1)\varphi(t)+(\lambda_1-\lambda_2)\varphi^2(t)-\lambda_2\varphi'(t). 
\ea \ee 
Having (\ref{82}) and  (\ref{83}), equations  (\ref{76})--(\ref{77})
can be rewritten as ODEs  for the functions $\varphi(t)$ and $
\psi(t)$:
\be\label{85}\varphi'(t)=-\frac{1}{(\lambda_1-\lambda_2)^2}\big(a_2-a_1+(\lambda_1-\lambda_2)
(\varphi+\psi)\big)\big((3\lambda_1-\lambda_2)\varphi+2\lambda_2\psi\big)\ee
\be\label{86}\psi'(t)=\frac{1}{(\lambda_1-\lambda_2)^2}\big(a_2-a_1+(\lambda_1-\lambda_2)
(\varphi+\psi)\big)\big(2\lambda_1\varphi+(3\lambda_2-\lambda_1)\psi\big).\ee

 Finally, using formulae (\ref{82})--(\ref{86}), the last two equations, (\ref{80})--(\ref{81}),
 can be rewritten as two algebraic equations to find $\varphi(t)$ and $
 \psi(t)$. The difference of those leads to  the classification equation
\be\label{87}\ba{l} \medskip
\Big(a_1-a_2-(\lambda_1-\lambda_2)(\varphi+\psi)\Big)(\lambda_1\varphi+\lambda_2\psi)
\Big(a_1(4\lambda_1+5\lambda_2)-a_2(5\lambda_1+4\lambda_2)-\\
-4(\lambda_1-\lambda_2)((2\lambda_1+\lambda_2)\varphi
+(\lambda_1+2\lambda\lambda_2)\psi)\Big)=0. 
\ea \ee 
 Thus, three subcases follow from (\ref{87}):

\medskip
1.3.1 $\varphi=-\psi+\frac{a_1-a_2}{\lambda_1-\lambda_2};$

\medskip
1.3.2 $\varphi=-\frac{\lambda_2}{\lambda_1}\psi;$

\medskip
1.3.3
$\varphi=\frac{1}{4(\lambda_1-\lambda_2)(2\lambda_1+\lambda_2)}\Big((4a_1-5a_2)\lambda_1
+(5a_1+4a_2)\lambda_2-4(\lambda_1-\lambda_2)(\lambda_1+2\lambda_2)\psi\Big).$
\medskip

In subcase  1.3.1, both equations, (\ref{80}) and  (\ref{81}), are
equivalent to the equation  \be\label{89}
\Big(a_1-(\lambda_1-\lambda_2)\psi\Big)\psi\Big(a_1-a_2-(\lambda_1-\lambda_2)\psi\Big)=0.
\ee \medskip If $\psi=\frac{a_1}{\lambda_1-\lambda_2},$ then
$\varphi=-\frac{a_2}{\lambda_1-\lambda_2}$ and  using (\ref{82}) and
(\ref{83})  we arrive at  operator (\ref{39}).\\\medskip If
$\psi=0$, then  $\varphi=\frac{a_1-a_2}{\lambda_1-\lambda_2} \ $, so
that operator  (\ref{40}) is obtained( the restriction $a_1\neq a_2$ guarantees that it is  no  Lie's operator ).\\
If  $\psi=\frac{a_1-a_2}{\lambda_1-\lambda_2} \ $, then $\varphi=0$,
what leads to the   operator $Q_3$ (\ref{40***}) and the same
restriction $a_1\neq a_2$. Thus, the proof of item {\it (i)} is
completed.

In subcase  1.3.2, both equations, (\ref{80}) and  (\ref{81}), are
equivalent to the equation
\be\label{88}
\psi(t)(a_1-a_2)(a_1\lambda_2-a_2\lambda_1)=0. \ee

Since  $\psi(t)\neq0$ (otherwise one arrives at the Lie operator
$Q=\p_t$) two possibilities occur: $a_1=a_2$ and
$a_1=\frac{\lambda_1}{\lambda_2}a_2$.

If $a_1=a_2$, then we obtain \[
\psi(t)=\frac{\lambda_1}{(\lambda_1-\lambda_2)t+\lambda_1\alpha}, \
\alpha=const \] from (\ref{86}).
  Note we  set  $\alpha=0$ without losing generality.
  Now  the functions  $p^1, \ p^2,$ can be found from
(\ref{83}), hence, we obtain operator  (\ref{42}). Operator
(\ref{39*}) follows from (\ref{38}) if one sets   $a_1=a_2=a$. Thus,
the proof of item {\it (ii)} is completed.

 If
$a_1=\frac{\lambda_1}{\lambda_2}a_2$ (here $a_2\not=0$ otherwise
item {\it (ii)} is obtained) then equation (\ref{86}) produces
\[\psi(t)=-\frac{\alpha_0a_2\lambda_1}{\exp(-\frac{a_2}{\lambda_2}t)-\alpha_0(\lambda_1-
\lambda_2)\lambda_2}, \] where $\alpha_0$ is a non-vanish constant.
 Thus, using equations (\ref{83}) and notations $a_2=a\lambda_2, \
\lambda_2 \alpha_0=\alpha,$ we obtain the most complicated operator
(\ref{44}). Finally,  operators  (\ref{39**}), (\ref{40*}) and
(\ref{40**}) are nothing else but those (\ref{39}), (\ref{40}) and
(\ref{40***}) with $a_1=a\lambda_1, a_2=a\lambda_2$, respectively.
Thus, all operators arising in item {\it (iii)} are constructed.

 It
turns out that the detailed analysis of subcase 1.3.3 doesn't lead
to any new operators.

The proof is now completed. \hfill $\blacksquare$

\textbf{Remark 1.} If  the restrictions $q^1_x=q^2_x=0$
 don't  take place we were not able to solve the corresponding
 nonlinear system of  DEs, hence, DLV system (\ref{2}) may
 admit $Q$-conditional
 symmetries of the form (\ref{19}) with $q^1_x \not =0$ and/or $q^2_x \not =0$.

\bt In the case  $\lambda_1=\lambda_2$,  DLV system (\ref{2}) admits
only such operators of the form (\ref{19}), which are equivalent to
the Lie symmetry operators.
  \et

\bt In the case  $\lambda_1\neq\lambda_2$,    DLV system (\ref{2})
is invariant under $Q$-conditional operators of the first type only
in two cases.  The corresponding systems   and $Q$-conditional
symmetries (up to local transformations $u\rightarrow bu, \
v\rightarrow \exp(\frac{a_2}{\lambda_2}t)v, \ b\neq0$  and
$u\rightarrow \exp(\frac{a_1}{\lambda_1}t)v,\ cv\rightarrow u, \
c\neq0$) have the forms \be\label{3-8}(i) \ba{l}
 \quad \ \lbd_1 u_t = u_{xx}+u(a_1+u+v),\\
 \quad \ \lbd_2 v_t = v_{xx}+ v(a_2+u+v),\quad a_1\neq a_2, \\
\ea \ee
\be\label{4-0}\quad{Q_1}=(\lbd_1-\lbd_2)\p_t+(a_1-a_2)u(\p_u-\p_v),
 \ee \be\label{4-1}\quad
{Q}_2=(\lbd_1-\lbd_2)\p_t-(a_1-a_2)v(\p_u - \p_v).\ee
\be\label{3-9}(ii) \ba{l}
 \quad \ \lbd_1 u_t = u_{xx}+u(a_1+u),\\
 \quad \ \lbd_2 v_t = v_{xx}+ v u, \\
\ea \ee \be \label{4-2} \quad \ \
{Q}=\p_t+\frac{2\al_1}{\lbd_1-\lbd_2} \, \p_x+\Big(\exp
(\al_1x+\frac{\al^2_1}{\lbd_2}t)\big((\al_3+\al_4\exp(-\frac{a_1}{\lbd_2}t))u+\al_3a_1\big)+\al_2v\Big)\p_v,
\ee where  $\alpha_k, \ k=
1,\dots,4$
 are arbitrary constants with the restriction
$\alpha^2_3+\alpha^2_4\neq0.$ There are no any other $Q$-conditional
operators of the first type.

In the case  $\lambda_1=\lambda_2$,    DLV system (\ref{2}) is
invariant  only under such $Q$-conditional operators of the first
type, which coincide with the Lie symmetry operators.
  \et

\textbf{Proofs} of Theorems 2 and 3 are  similar to one presented
above for Theorem 1 and omitted here because their bulk. It should
be noted that  both manifolds arising in Definition 1  and the most
general form (\ref{3*}) of the operator in question   were  used.

\textbf{Remark 2.} Theorems 2 and 3 give {\it a complete description
of $Q$-conditional symmetries of the first type in explicit form}
because
 there aren't any additional restrictions on the form  of those
operators (in contrary to the  $Q$-conditional symmetries of the
second type).

\newpage

  \begin{center}
{\textbf{4. Reductions to ODEs' systems, exact solutions and their
application}}
\end{center}

First of all, we note that DLV system  (\ref{2}) is invariant under
time and space translations, hence,   its arbitrary  solution
$u_0(t,x), \ v_0(t,x)$ generates a two-parameter family of solutions
of the form  $u_0(t-t_0,x-x_0), \ v_0(t-t_0,x-x_0).$ Having this in
mind, we set $t_0=x_0=0$ in the solutions obtained below.

It is well-known that using $Q$-conditional symmetries one can
reduce the given two-dimensional PDE (system of PDEs) to an ODE
(system of ODEs) via  the same procedure as for classical Lie
symmetries. Thus, to construct an ansatz corresponding to the
operator $Q$, the system of the linear first-order PDEs
 \be\label{92*} \ba{l}
{Q}u=0,\\
{Q}v=0 \ea \ee should be solved. Substituting the ansatz obtained
into DLV system  with correctly-specified coefficients, one obtains
an system of ODEs, i.e., the reduced system of equations. Since this
procedure is the same for all operators, we consider  only operator
(\ref{39}) in details. In this case system (\ref{92*}) takes the
form
\be\label{93} \ba{l} 
(\lambda_1-\lambda_2)u_t=-(a_1v+a_2u+a_1a_2),\\
(\lambda_1-\lambda_2)v_t=a_1v+a_2u+a_1a_2.
\ea \ee 
To solve (\ref{93}) we immediately note that  $u_t=-v_t$, hence,
\be\label{94}u(t,x)=-v(t,x)+\varphi_1(x).\ee Substituting (\ref{94})
into the second equation of  (\ref{93}), we arrive at the equation
\[(\lambda_1-\lambda_2)v_t=(a_1-a_2)v+a_2\varphi_1(x)+a_1a_2.\]
If  $a_1\neq a_2,$ then this equation has the general solution
\[v(t,x)=\frac{1}{a_1-a_2}\big(\exp(\frac{a_1-a_2}
{\lambda_1-\lambda_2}t)\varphi_2(x)-a_2\varphi_1(x)-a_1a_2\big),\]
therefore the ansatz
\be\label{95} \ba{l} \medskip
u(t,x)=\frac{1}{a_1-a_2}\big(-\exp(\frac{a_1-a_2}
{\lambda_1-\lambda_2}t)\varphi_2(x)+a_1\varphi_1(x)+a_1a_2\big),\\
v(t,x)=\frac{1}{a_1-a_2}\big(\exp(\frac{a_1-a_2}
{\lambda_1-\lambda_2}t)\varphi_2(x)-a_2\varphi_1(x)-a_1a_2\big)
\ea \ee is obtained. Here  $\varphi_1$ and $\varphi_2$ are to be
found functions.

If  $a_1=a_2=a,$ then the ansatz
\be\label{96} \ba{l} \medskip
u(t,x)=\varphi_1(x)-\varphi_2(x)-\frac{a}{\lambda_1-\lambda_2}(\varphi_1(x)+a)t,\\
v(t,x)=\varphi_2(x)+\frac{a}{\lambda_1-\lambda_2}(\varphi_1(x)+a)t
\ea \ee  is obtained.

To construct the reduced system, we substitute ansatz  (\ref{95})
into (\ref{38}). It means that we simply calculate the derivatives
$u_t, \ v_t, \ u_{xx}, \ v_{xx},$  and insert them into (\ref{38}).
After the relevant simplifications one arrives at  the ODEs system
\be\label{98} \ba{l} \medskip
\varphi''_1+\varphi^2_1+(a_1+a_2)\varphi_1+a_1a_2=0,\\
\varphi''_2+\frac{a_2\lambda_1-a_1\lambda_2}{\lambda_1-\lambda_2}\varphi_2+
\varphi_1\varphi_2=0
\ea \ee 
to find the functions $\varphi_1$ and $\varphi_2$. Similarly, ansatz
(\ref{96}) leads to the reduced system of equations \be\label{98***}
\ba{l} \medskip
\varphi''_1+(a+\varphi_1)^2=0,\\
\varphi''_2+(\varphi_2-\frac{a\lambda_2}{\lambda_1-\lambda_2})
(a+\varphi_1)=0. \ea \ee

In a quite similar way other operators listed in Theorem 1 were used
to find ans\"atze and reduced systems of ODEs. They are  presented
in Table 1.

Now we construct exact solutions of DLV system using the ans\"atze
and the reduced systems obtained above. It should be stressed that
all the ODE systems listed in Table 1 are nonlinear and none of them
can be easily integrated.

 Let us consider system (\ref{98})
obtained by application of ansatz (\ref{95}). Since the general
solution of this nonlinear ODE systems cannot be found in an
explicit form, we look for particular solutions. Setting
$\varphi_1=\alpha=const,$  we find
\[ \alpha^2+(a_1+a_2)\alpha+a_1a_2=0 \ \Rightarrow \alpha_1=-a_1, \
\alpha_2=-a_2 \] from the first equation of  system (\ref{98}).
 Now we take  $\varphi_1=-a_1$ (the case $\varphi_1=-a_2$ leads
 to the solution with the same structure) and substitute into the second
 equation of system (\ref{98}):
 \be\label{105}
 \varphi''_2-\beta \lambda_1\varphi_2=0,
 \ee
where $\beta=\frac{a_1-a_2}{\lambda_1-\lambda_2}\not=0.$ Depending
on sign of the parameter  $\beta$ the linear ODE (\ref{105})
generates two families of the general solutions. Using those
solutions and ansatz (\ref{95}), we obtain  the following two
families of exact solutions of the DLV system (\ref{38}):
 \be\label{106} \ba{l}
\medskip
u(t,x)=-a_1+\frac{1}{a_2-a_1}\big(C_1\exp(\sqrt{\beta\lambda_1}x)+C_2\exp(-\sqrt{\beta\lambda_1}x)\big)e^{\beta t},\\
v(t,x)=\frac{1}{a_1-a_2}\big(C_1\exp(\sqrt{\beta\lambda_1}x)+C_2\exp(-\sqrt{\beta\lambda_1}x)\big)e^{\beta
t}, \ea \ee if $\beta>0,$  and   \be\label{107} \ba{l}
\medskip
u(t,x)=-a_1+\frac{1}{a_2-a_1}\big(C_1\cos(\sqrt{-\beta\lambda_1}x)+C_2\sin(\sqrt{-\beta\lambda_1}x)\big)e^{\beta t},\\
v(t,x)=\frac{1}{a_1-a_2}\big(C_1\cos(\sqrt{-\beta\lambda_1}x)+C_2\sin(\sqrt{-\beta\lambda_1}x)\big)e^{\beta
t}, \ea \ee if  $\beta<0$ (hereafter  $C_1,  C_2$ are arbitrary
constants).

\newpage

{\bf Table 1.  Ans\"atze and reduced systems of ODEs for DLV system
 (\ref{2})}
\begin{small}
\begin{center}
\begin{tabular}{|c|c|c|c|
} \hline

  &$Q_i$ &   Ans\"atze  & Systems of ODEs    \\

\hline &&&\\ 1 & (\ref{37})&$u(t,x)=\varphi_1(\omega), \
\omega=x-C_1t$&$\varphi''_1+C_1\lambda_1\varphi'_1+(a_1+\varphi_1)\varphi_1=0$
\\
 & &$v(t,x)=\varphi_2(\omega)e^{C_2t}+
\exp\big(\frac{\lambda_1-\lambda_2}{2}C_1\omega+At\big)\times$ &
$\varphi''_2+
C_1\lambda_2\varphi'_2+\varphi_2(\varphi_1-C_2\lambda_2)=0$\\
 & & $ \Big((C_3+C_4\exp(\frac{a_1}{\lambda_2}t))\varphi_1(\omega)
+a_1C_4\exp(\frac{a_1}{\lambda_2}t)\Big)$&\\
\hline
 &&& \\

2&(\ref{39})&$u(t,x)=\frac{1}{a_1-a_2}\big(-\exp(\frac{a_1-a_2}{\lambda_1-\lambda_2}t)\varphi_2(x)
+$&$   \varphi''_1+\varphi^2_1+(a_1+a_2)\varphi_1+a_1a_2=0 $\\
& &
$a_1\varphi_1(x)+a_1a_2\big)$&\\
& &
$v(t,x)=\frac{1}{a_1-a_2}\big(\exp(\frac{a_1-a_2}{\lambda_1-\lambda_2}t)\varphi_2(x)
 -$&$\varphi''_2+\frac{a_2\lambda_1-a_1\lambda_2}{\lambda_1-\lambda_2}\varphi_2+
\varphi_1\varphi_2=0$\\
& &
$a_2\varphi_1(x)-a_1a_2\big)$&$ $\\
\hline
 &&& \\
3&
(\ref{40})&$u(t,x)=\varphi_2(x)\exp(\frac{a_1-a_2}{\lambda_1-\lambda_2}t)$&$\varphi''_1+\varphi^2_1+a_2\varphi_1=0$\\
& &
$v(t,x)=\varphi_1(x)-\varphi_2(x)\exp(\frac{a_1-a_2}{\lambda_1-\lambda_2}t)$&$\varphi''_2+\frac{a_2\lambda_1-a_1\lambda_2}{\lambda_1-\lambda_2}\varphi_2+
\varphi_1\varphi_2=0$\\
\hline
 &&& \\
4&
(\ref{40***})&$u(t,x)=\varphi_1(x)-\varphi_2(x)\exp(\frac{a_1-a_2}{\lambda_1-\lambda_2}t)$&$\varphi''_1+\varphi^2_1+a_1\varphi_1=0$\\
& &
$v(t,x)=\varphi_2(x)\exp(\frac{a_1-a_2}{\lambda_1-\lambda_2}t)$&$\varphi''_2+\frac{a_2\lambda_1-a_1\lambda_2}{\lambda_1-\lambda_2}\varphi_2+
\varphi_1\varphi_2=0$\\
\hline
 &&& \\
5&
(\ref{39*})&$u(t,x)=\varphi_1(x)-\varphi_2(x)-\frac{a}{\lambda_1-\lambda_2}(\varphi_1(x)+a)t$&$\varphi''_1+(a+\varphi_1)^2=0$\\
& &$v(t,x)=\varphi_2(x)+\frac{a}{\lambda_1-\lambda_2}(\varphi_1(x)
+a)t$&$\varphi''_2+(\varphi_2-\frac{a\lambda2}{\lambda_1-\lambda_2})(a+\varphi_1)=0$\\
\hline
 &&& \\
6&
(\ref{42})&$u(t,x)=\frac{1}{\lambda_1-\lambda_2}\big(\lambda_1\varphi_1(x)-
t\varphi_2(x)\big)$&$\varphi''_1+\varphi_2+\varphi_1(a+\varphi_1)=0$\\
& &
$v(t,x)=\frac{1}{\lambda_1-\lambda_2}\big(-\lambda_2\varphi_1(x)+
t\varphi_2(x)\big)$&$\varphi''_2+\varphi_2(a+\varphi_1)=0$\\
\hline
 &&& \\
7&
(\ref{39**})&$u(t,x)=\frac{1}{a(\lambda_1-\lambda_2)}\big(-e^{at}\varphi_2(x)
+$& $  \varphi''_1+\varphi^2_1+a(\lambda_1+\lambda_2)\varphi_1
+a^2\lambda_1\lambda_2=0 $\\
& &
$a\lambda_1\varphi_1(x)-a^2\lambda_1\lambda_2\big)$&$$\\
 & &
$v(t,x)=\frac{1}{a(\lambda_1-\lambda_2)}\big(e^{at}\varphi_2(x)
 -$&$\varphi''_2+
\varphi_1\varphi_2=0$\\
 & &
$a\lambda_2\varphi_1(x)-a^2\lambda_1\lambda_2\big)$&$$\\
 \hline
 &&& \\
8&
(\ref{40*})&$u(t,x)=e^{at}\varphi_2(x)$&$\varphi''_1+\varphi^2_1+a\lambda_2\varphi_1=0$\\
& &$v(t,x)=\varphi_1(x)-e^{at}\varphi_2(x)$&$\varphi''_2+
\varphi_1\varphi_2=0$\\
\hline
 &&& \\
9&
(\ref{40**})&$u(t,x)=\varphi_1(x)-e^{at}\varphi_2(x)$&$\varphi''_1+\varphi^2_1+a\lambda_1\varphi_1=0$\\
& & $v(t,x)=e^{at}\varphi_2(x)$&$\varphi''_2+
\varphi_1\varphi_2=0$\\
\hline
 &&& \\
10 &
(\ref{44})&$u(t,x)=\frac{1}{\lambda_1-\lambda_2}\big(\lambda_1\varphi_1(x)+
a\lambda_1\lambda_2-$&
$\varphi''_1+\varphi^2_1-a\varphi_2+$\\
& &$\varphi_2(x)(1-\alpha(\lambda_1-\lambda_2) e^{at})\big)$&$a(\lambda_1+\lambda_2)\varphi_1+a^2\lambda_1\lambda_2=0$\\
& &
$v(t,x)=\frac{1}{\lambda_1-\lambda_2}\big(\varphi_2(x)(1-\alpha(\lambda_1-\lambda_2) e^{at})-$&$\varphi''_2+\varphi_1\varphi_2=0$\\
& &$\lambda_2\varphi_1(x)- a\lambda_1\lambda_2\big)$&$$\\

 \hline
\end{tabular}
\end{center}
\end{small}

\textbf{ Remark 3.} In Table 1, the parameter
$A=\frac{\lambda^2_1-\lambda^2_2}{4\lambda_2}C^2_1-\frac{a_1}{\lambda_2},$
while  $C_k, \ k=
1,\dots,4$ are arbitrary constants.

\vspace{0.5cm}

Let construct solutions of  (\ref{98}) with some restrictions on
$\lambda_1$ and $\lambda_1$. Firstly, we note that the substitution
\be\label{108} \varphi_1=\varphi-a_1 \ee simplifies the first
equation of  (\ref{98}) to the form  \be\label{109}
\varphi''+\varphi^2+(a_2-a_1)\varphi=0. \ee Of course, (\ref{109})
can be reduced to the first-order ODE
\be\label{110}\Big(\frac{d\varphi}{dx}\Big)^2=-\frac{2}{3} \
\varphi^3+(a_1-a_2)\varphi^2+C \ \ee with the general solution
containing special functions, Weierstrass functions    \cite{b-e}.
To avoid cumbersome formulae, we set   $C=0$, hence, the general
solution  is \be\label{111}
\varphi=\frac{3}{2}(a_1-a_2)\big(1-\tanh^2(\frac{1}{2}\sqrt{a_1-a_2}
\,x)\big) ,\ee \be\label{112}
\varphi=\frac{3}{2}(a_1-a_2)\big(1-\coth^2(\frac{1}{2}\sqrt{a_1-a_2}\,x)\big)
,\ee if  $a_1>a_2$,  and \be\label{113}
\varphi=\frac{3}{2}(a_1-a_2)\big(1+\tan^2(\frac{1}{2}\sqrt{a_2-a_1}\,x)\big)
,\ee   if $a_1<a_2.$

Thus, we can apply each of formulae (\ref{111})--(\ref{113}) to
solve the second equation of (\ref{98}). In the case of solution
(\ref{111}), this ODE takes the form  \be\label{114}
\varphi_2''+\varphi_2(a_1-a_2)\Big(\frac{\lambda_1-3\lambda_2}{2(\lambda_1-\lambda_2)}-
\frac{3}{2}\tanh^2(\frac{1}{2}\sqrt{a_1-a_2}\,x)\Big)=0 .\ee
Nevertheless, the general solution of (\ref{114}) is still unknown,
its  particular solutions can be found \cite{pol-za}:
 \be\label{115} \varphi_2=
\cosh^3(\frac{1}{2}\sqrt{a_1-a_2}\,x),\ee if
$\lambda_1=\frac{9}{5}\lambda_2$, and
 \be\label{117}
\varphi_2=\sinh(\frac{1}{2}\sqrt{a_1-a_2}\,x)
\cosh^3(\frac{1}{2}\sqrt{a_1-a_2}\,x),\ee if
$\lambda_1=\frac{4}{3}\lambda_2$.

 Thus, substituting  the functions $\varphi_1(x)$ and $\varphi_2(x)$
given  by  formulae (\ref{108}), (\ref{111}) and (\ref{115}) into
ansatz (\ref{95}), we find the exact solution of DLV system
(\ref{38})
\be\label{116} \ba{l} \medskip
u(t,x)=\frac{1}{2}a_1-\frac{3}{2}a_1\tanh^2(\frac{1}{2}\sqrt{a_1-a_2}\,x)-\\\medskip
\quad \quad \quad  -\frac{1}{a_1-a_2}\cosh^3(\frac{1}{2}\sqrt{a_1-a_2}\,x)\exp(\frac{5(a_1-a_2)}{4\lambda_2}t),\\\medskip
v(t,x)=-\frac{3}{2}a_2(1-\tanh^2(\frac{1}{2}\sqrt{a_1-a_2}\,x))+\\
\quad \quad \quad  +\frac{1}{a_1-a_2}\cosh^3(\frac{1}{2}\sqrt{a_1-a_2}\,x)\exp(\frac{5(a_1-a_2)}{4\lambda_2}t).
\ea \ee if $\lambda_1=\frac{9}{5}\lambda_2$, $a_1>a_2$.

Similarly, solution (\ref{117}) leads to the exact solution
\be\label{118} \ba{l} \medskip
u(t,x)=\frac{1}{2}a_1-\frac{3}{2}a_1\tanh^2(\frac{1}{2}\sqrt{a_1-a_2}\,x)-\\\medskip
\quad \quad \quad
-\frac{1}{a_1-a_2}\sinh(\frac{1}{2}\sqrt{a_1-a_2}x)\cosh^3(\frac{1}{2}\sqrt{a_1-
a_2}\,x)\exp(\frac{3(a_1-a_2)}{\lambda_2}t),\\\medskip
v(t,x)=-\frac{3}{2}a_2(1-\tanh^2(\frac{1}{2}\sqrt{a_1-a_2}\,x))+\\
\quad \quad \quad
+\frac{1}{a_1-a_2}\sinh(\frac{1}{2}\sqrt{a_1-a_2}\,x)\cosh^3(\frac{1}{2}\sqrt{a_1
-a_2}\,x)\exp(\frac{3(a_1-a_2)}{\lambda_2}t)
\ea \ee  of DLV system (\ref{38}) with
$\lambda_1=\frac{4}{3}\lambda_2$, $a_1>a_2$.

In a quite similar way solutions  (\ref{112}) and  (\ref{113}) were
also  used to construct three new    solutions of  DLV system
(\ref{38}). Omitting  straightforward calculations we present only
the result:
\be\label{119} \ba{l} \medskip
u(t,x)=\frac{1}{2}a_1-\frac{3}{2}a_1\coth^2(\frac{1}{2}\sqrt{a_1-a_2}\,x)-\\\medskip
\quad \quad \quad  -\frac{1}{a_1-a_2}\sinh^3(\frac{1}{2}\sqrt{a_1-a_2}\,x)\exp(\frac{5(a_1-a_2)}{4\lambda_2}t),\\\medskip
v(t,x)=-\frac{3}{2}a_2(1-\coth^2(\frac{1}{2}\sqrt{a_1-a_2}\,x))+\\\medskip
\quad \quad \quad  +\frac{1}{a_1-a_2}\sinh^3(\frac{1}{2}\sqrt{a_1-a_2}\,x)\exp(\frac{5(a_1-a_2)}{4\lambda_2}t),
\ea \ee 
if  $\lambda_1=\frac{9}{5}\lambda_2$, $a_1>a_2;$
\be\label{120} \ba{l} \medskip
u(t,x)=\frac{1}{2}a_1-\frac{3}{2}a_1\coth^2(\frac{1}{2}\sqrt{a_1-a_2}\,x)-\\\medskip
\quad \quad \quad
-\frac{1}{a_1-a_2}\cosh(\frac{1}{2}\sqrt{a_1-a_2}\,x)\sinh^3(\frac{1}{2}\sqrt{a_1-
a_2}\,x)\exp(\frac{3(a_1-a_2)}{\lambda_2}t),\\\medskip
v(t,x)=-\frac{3}{2}a_2(1-\coth^2(\frac{1}{2}\sqrt{a_1-a_2}\,x))+\\\medskip
\quad \quad \quad
+\frac{1}{a_1-a_2}\cosh(\frac{1}{2}\sqrt{a_1-a_2}\,x)\sinh^3(\frac{1}{2}\sqrt{a_1
-a_2}\,x)\exp(\frac{3(a_1-a_2)}{\lambda_2}t),
\ea \ee 
if $\lambda_1=\frac{4}{3}\lambda_2$, $a_1>a_2;$
\be\label{121} \ba{l} \medskip
u(t,x)=\frac{1}{2}a_1+\frac{3}{2}a_1\tan^2(\frac{1}{2}\sqrt{a_2-a_1}\,x)-\\\medskip
\quad \quad \quad  -\frac{1}{a_1-a_2}\cos^3(\frac{1}{2}\sqrt{a_2-a_1}\,x)\exp(\frac{5(a_1-a_2)}{4\lambda_2}t),\\\medskip
v(t,x)=-\frac{3}{2}a_2(1+\tan^2(\frac{1}{2}\sqrt{a_2-a_1}\,x))+\\\medskip
\quad \quad \quad  +\frac{1}{a_1-a_2}\cos^3(\frac{1}{2}\sqrt{a_2-a_1}\,x)\exp(\frac{5(a_1-a_2)}{4\lambda_2}t),
\ea \ee 
if $\lambda_1=\frac{9}{5}\lambda_2$, $a_1<a_2$.

\bigskip

In a  similar way one may use other ans\"atze and reduced systems of
ODEs for constructing exact solutions of DLV system (\ref{2}) with
the correctly-specified coefficients. Let us consider the most
cumbersome ansatz and reduced system, which correspond to the
$Q$-conditional operator (\ref{44}). Nevertheless  the reduced
system of ODEs (see  case 10 of Table 1) is again non-integrable,
its particular solutions can be derived by setting
 \be\label{122}
\varphi_2=\lambda_1 \varphi_1+a\lambda_1\lambda_2.\ee The reduced
system under condition  (\ref{122}) takes the form
 \be\label{123} \varphi_1''+\varphi^2_1+a\lambda_2\varphi_1=0. \ee
Since ODE (\ref{123}) has the same structure as
 (\ref{109}), we immediately obtain its solutions
 (\ref{111})--(\ref{113}) with $a_2-a_1=a\lambda_2$.
Thus, substituting (\ref{122}) and (\ref{111})--(\ref{113}) with
$a_2-a_1=a\lambda_2$ into the  ansatz listed in the  last case of
Table 1, we find the exact solutions of  DLV system (\ref{43})
\be\label{127} \ba{l}
\medskip
u(t,x)=\alpha
a\lambda_1\lambda_2\big(-\frac{1}{2}+\frac{3}{2}\tanh^2(\frac{1}{2}\sqrt{-a\lambda_2}\,x)\big)e^{at}
,\\
\medskip%
v(t,x)=-\frac{3}{2}a\lambda_2\big(1-\tanh^2(\frac{1}{2}\sqrt{-a\lambda_2}\,x)\big)-\\
\qquad \qquad -\alpha a\lambda_1\lambda_2(-\frac{1}{2}+\frac{3}{2}\tanh^2\big(\frac{1}{2}\sqrt{-a\lambda_2}\,x)\big)e^{at},
\ea \ee 
\be\label{128} \ba{l} \medskip
u(t,x)=\alpha
a\lambda_1\lambda_2\big(-\frac{1}{2}+\frac{3}{2}\coth^2(\frac{1}{2}\sqrt{-a\lambda_2}\,x)\big)e^{at}
,\\\medskip
v(t,x)=-\frac{3}{2}a\lambda_2\big(1-\coth^2(\frac{1}{2}\sqrt{-a\lambda_2}x)\big)-\\ \medskip
\qquad \quad \quad -\alpha a\lambda_1\lambda_2\big(-\frac{1}{2}+\frac{3}{2}\coth^2(\frac{1}{2}\sqrt{-a\lambda_2}\,x)\big)e^{at},
\ea \ee 
if $a<0$  and
\be\label{129} \ba{l} \medskip
u(t,x)=-\alpha
a\lambda_1\lambda_2\big(\frac{1}{2}+\frac{3}{2}\tan^2(\frac{1}{2}\sqrt{a\lambda_2}\,x)\big)e^{at}
,\\
v(t,x)=-\frac{3}{2}a\lambda_2\big(1+\tan^2(\frac{1}{2}\sqrt{a\lambda_2}\,x)\big) +
\alpha a\lambda_1\lambda_2\big(\frac{1}{2}+\frac{3}{2}\tan^2(\frac{1}{2}\sqrt{a\lambda_2}\,x)\big)e^{at},
\ea \ee 
if $a>0.$

It should be noted that all the solutions obtained above cannot be
constructed using Lie symmetries. In fact DLV systems (\ref{38}) and
(\ref{43}) don't admit any non-trivial Lie symmetry, hence, the
standard plane wave solutions only  can be found by Lie symmetry
reductions.

Finally, we present an example that demonstrates remarkable
properties of some solutions presented above.

\begin{figure}[t]\label{Fig-1}
\begin{minipage}[t]{8cm}
  \quad  \quad \centerline{\includegraphics[width=9cm]{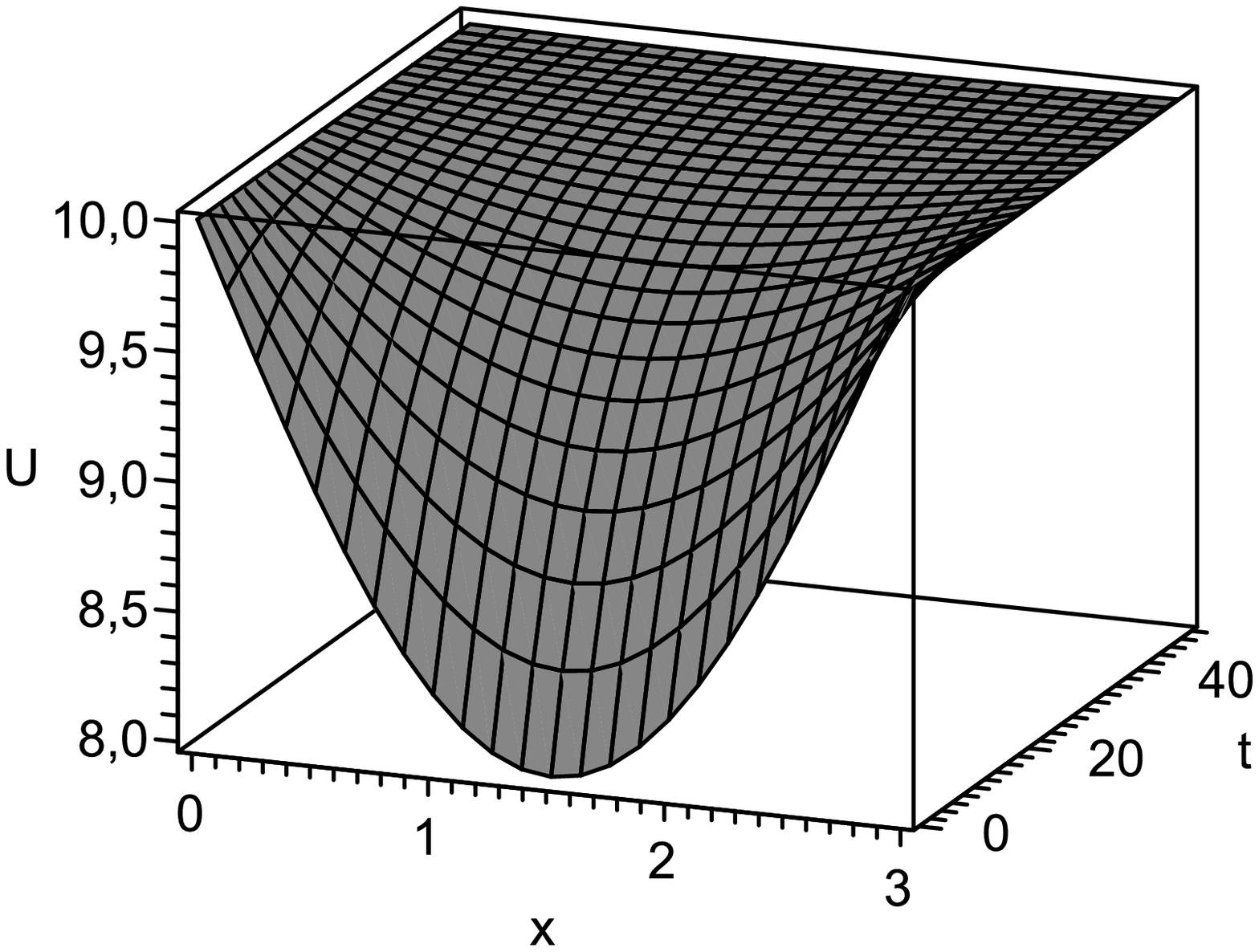}}
\end{minipage}
\hfill
\begin{minipage}[t]{8cm}
\centerline{\includegraphics[width=9cm]{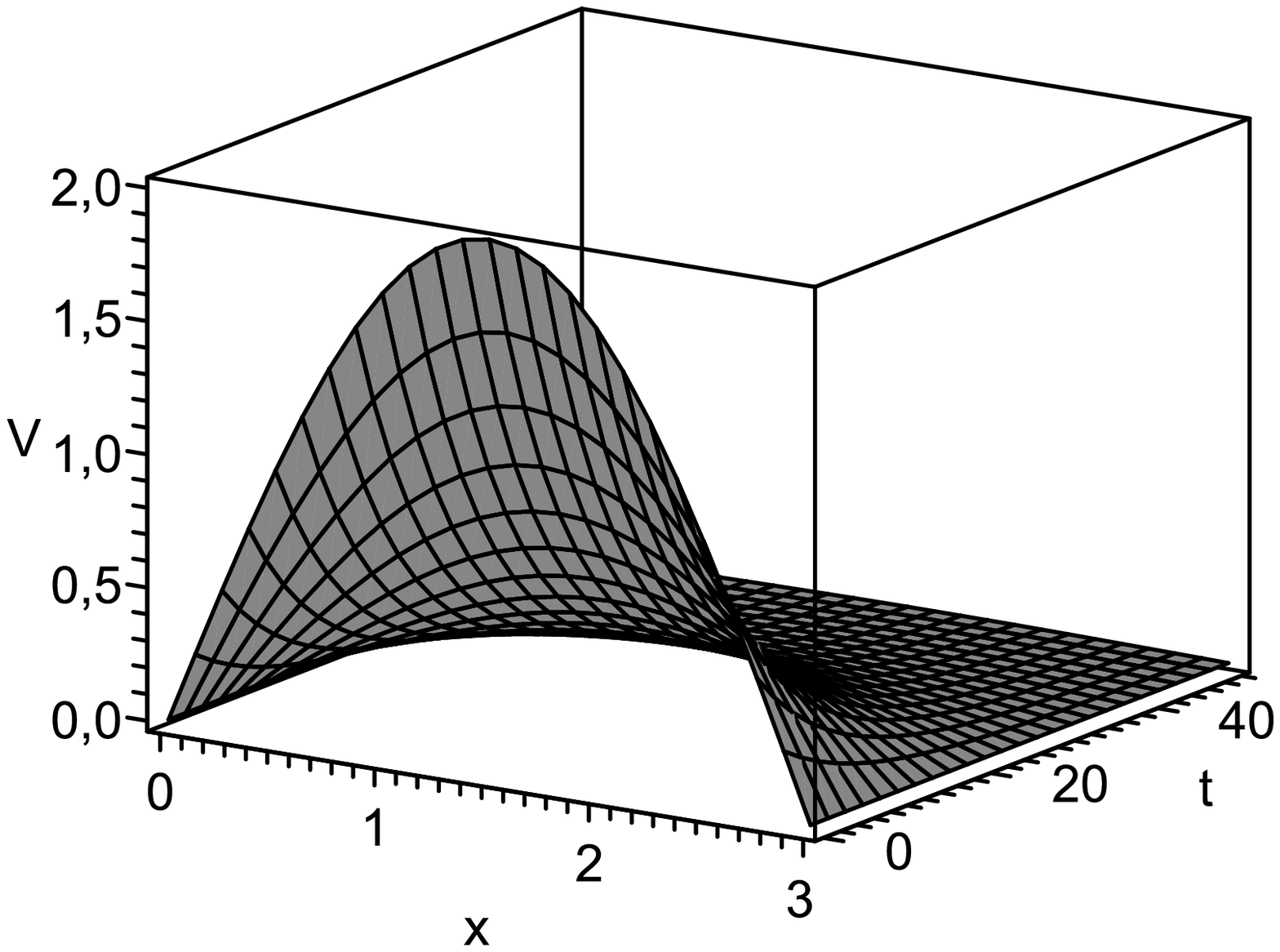}}
\end{minipage}
\center\caption{ Solution  (\ref{134}) of system (\ref{136}) with
$a_1=1,\ a_2=2,\ \lambda_1=11,\ \lambda_2=1, \ b=0.1, \ c=0.1, \ \
C_2=0.2, \ \beta=~-0.1.$}
\end{figure}

\vspace{0.5cm}

\textbf{Example.} Consider solution  (\ref{107}) with  $C_1=0$.
 Using substitution  $u=-bU,\ v=-cV \ (b>0,\ c>0),$
 one transforms DLV system  (\ref{38}) to the system describing the competition of
two species
 \be\label{136} \ba{l}
 \lbd_1 U_t = U_{xx}+U(a_1-bU-cV),\\
 \lbd_2 V_t = V_{xx}+ V(a_2-bU-cV)\ea \ee
and   solution  (\ref{107}) to the form
\be\label{134} \ba{l}
\medskip
U(t,x)=\frac{a_1}{b}+\frac{1}{(a_1-a_2)b} \ C_2\sin(\sqrt{-\beta\lambda_1}x)e^{\beta t},\\
V(t,x)=\frac{1}{(a_2-a_1)c} \
C_2\sin(\sqrt{-\beta\lambda_1}x)e^{\beta t}, \ea \ee where the
coefficient restrictions $\beta \equiv
\frac{a_1-a_2}{\lambda_1-\lambda_2}<0, \ a_1>0,\ a_2>0$ are assumed.
Using this solution one may formulate the following  theorem giving
the classical solution for the  BVP with the constant Dirichlet
conditions on the boundaries.

\bt  The classical  solution of boundary-value problem for the
competition  system (\ref{136})
 and the initial profile
 \be\label{134a} \ba{l}
\medskip
U(0,x)=\frac{a_1}{b}+\frac{1}{(a_1-a_2)b} \ C_2\sin(\sqrt{-\beta\lambda_1}x),\\
V(0,x)=\frac{1}{(a_2-a_1)c} \ C_2\sin(\sqrt{-\beta\lambda_1}x), \ea
\ee
 and boundary conditions
\be\label{134b} \ba{l} x=0:\  U= \frac{a_1}{b}, \ V=0,  \\
x=\frac{\pi}{\sqrt{-\beta\lambda_1}}:\  U= \frac{a_1}{b}, \ V=0, \ea
\ee
 in domain  $\Omega=\{ (t,x) \in (0,+ \infty )\times
\Bigl(0,\frac{\pi}{\sqrt{-\beta\lambda_1}}\Bigr)\} $
 is given by formulae (\ref{134}).
  \et

The solution  (\ref {134}) with $\beta < 0 $
 has the time asymptotic
 \be\label{135} (U,\ V)\rightarrow (\frac{a_1}{b}, \ 0), \quad
t\rightarrow +\infty.\ee
 Thus, this solution  describes  the
competition between the two species  when   the species $U$
eventually
dominate  while  the species $V$ die. An example of this competition
with the correctly-specified coefficients is presented on Fig.1.

 \begin{center}
{\textbf{5. Conclusions}}
\end{center}

In this paper  $Q$-conditional symmetries of
 the diffusive   Lotka-Volterra system (\ref {2}) and their application for finding exact solutions
 are studied.
Following the recent paper \cite{ch-2010}, two different definitions
of such symmetries are used to derive systems of DEs and to
construct them in the explicit form. It turns out that
$Q$-conditional operators of the first type can be derived much
easier than those of the second type (nonclassical symmetries),
hence, Theorem 3 was proved, which completes description of
$Q$-conditional operators of the first type. Nevertheless Theorem 1
 presents a wider list of $Q$-conditional operators
(because all the operators of the first type are automatically among
those of the second type), the result is still incomplete. In fact,
the additional restriction on the form of operators in question was
used.

All the $Q$-conditional operators obtained were used to construct
non-Lie ans\"atze and to reduce DLV system (\ref {2}) to the
corresponding systems of ODEs. As result,  a wide range of new exact
solutions in explicit form  was found. These  solutions possess a
complicated structure and cannot be found by the classical Lie
algorithm.  In the particular case, they  differs from the standard
plane wave solutions obtained in \cite{ch-du-04, rod-mimura-2000}.
Note plane wave solutions can be derived from some ans\"atze listed
in Table 1 under additional restrictions (see, e.g., ans\"atze with
$\varphi_1=const $ in 3-rd and 4-th cases of the table).

Finally,
 a  realistic  interpretation for competing species
 has been  provided for  exact solution (\ref {134}).
 In fact, this solution  describes  the
competition between  two populations of species  when   one of them
eventually leads to the extinction of other.

The work is in progress to construct $Q$-conditional symmetries and
exact solutions of   {\it a multicomponent} DLV system.

\end{document}